\documentclass[twocolumn]{autart}    

\usepackage{amsfonts,amssymb,mathtools} 
\usepackage{graphicx,subcaption} 
\usepackage[dvipsnames]{xcolor}   
\usepackage{algpseudocode}
\usepackage[chapter]{algorithm} 
\usepackage{tikz,pgfplots}
\usepgfplotslibrary{colormaps,fillbetween,patchplots}
\usetikzlibrary{shapes.multipart}
\usetikzlibrary[pgfplots.groupplots]
\newcommand{\ignore}[1]{}
\pgfplotsset{compat=newest,tick label style={font=\small},
	every axis/.append style={label style={font=\small}}}    

\newcommand{\nc}{n_{\mathrm{c}}}
\newcommand{\Ccal}{\mathcal{C}}
\newcommand{\Dcal}{\mathcal{D}}
\newcommand{\Hcal}{\mathcal{H}}
\newcommand{\Jcal}{\mathcal{J}}
\newcommand{\Kcal}{\mathcal{K}}
\newcommand{\Rcal}{\mathcal{R}}

\newcommand{\blue}[1]{{#1}}

\expandafter\def\expandafter\normalsize\expandafter{%
	\normalsize%
	\setlength\abovedisplayskip{7pt}%
	\setlength\belowdisplayskip{8pt}%
	\setlength\abovedisplayshortskip{-8pt}%
	\setlength\belowdisplayshortskip{2pt}%
}

\DeclareMathOperator{\dom}{dom}

\begin{document}

\begin{frontmatter}
\runtitle{CB-PWA Approximation of Nonlinear Systems} 

\title{Iterative Cut-Based PWA Approximation of Multi-Dimensional Nonlinear Systems\thanksref{footnoteinfo}}

\thanks[footnoteinfo]{Corresponding author L.~Gharavi.}

\author[dcsc]{Leila Gharavi}\ead{L.Gharavi@tudelft.nl},    
\author[dcsc]{Bart De Schutter}\ead{B.DeSchutter@tudelft.nl},              
\author[seuni]{Simone Baldi}\ead{103009004@seu.edu.cn}

\address[dcsc]{Delft Center for Systems and Control, Delft, The Netherlands}
\address[seuni]{Southeast University, Nanjing, China}             
        

\begin{abstract}  
		                        
PieceWise Affine (PWA) approximations for nonlinear functions have been extensively used for tractable, computationally efficient control of nonlinear systems. However, reaching a desired approximation accuracy without prior information about the behavior of the nonlinear systems remains a challenge in the function approximation and control literature. As the name suggests, PWA approximation aims at approximating a nonlinear function or system by dividing the domain into multiple subregions where the nonlinear function or dynamics is approximated locally by an affine function also called local mode. Without prior knowledge of the form of the nonlinearity, the required number of modes, the locations of the subregions, and the local approximations need to be optimized simultaneously, which becomes highly complex for large-scale systems with multi-dimensional nonlinear functions. This paper introduces a novel approach for PWA approximation of multi-dimensional nonlinear systems, utilizing a hinging hyperplane formalism for cut-based partitioning of the domain. The complexity of the PWA approximation is iteratively increased until reaching the desired accuracy level. Further, the tractable cut definitions allow for different forms of subregions, as well as the ability to impose continuity constraints on the PWA approximation. The methodology is explained via multiple examples and its performance is compared to two existing approaches through case studies, showcasing its efficacy.
\end{abstract}

\end{frontmatter}

\section{Introduction}

PWA systems are a class of hybrid modeling frameworks where the dynamics is expressed by multiple subsystems, i.e.\ local modes, that are affine functions of states and inputs and are active on a partition of the domain region, i.e.\ subregions~\cite{Lauer2019}. PWA approximation of nonlinear functions has long been used in diverse applications, contributing to enhanced modeling power~\cite{Gorokhovik1994}, improved computational efficiency~\cite{Bayat2012}, identification of explicit control laws~\cite{Gersnoviez2019} or serving as a descriptor for neural networks in machine learning~\cite{Bertsimas2019}. Moreover, complexity of PWA approximations~\cite{Lauer2015} and their verification processes~\cite{Brox2013} has been examined in the literature. 

Approximating a nonlinear function by a PWA form is rather straightforward if the subregions are known a~priori, as the problem boils down to determining local affine approximations on each subregion. The knowledge of the subregions may arise from the knowledge of different regimes (refer to the tire model in~\cite{Sun2022}) or from the knowledge of different equilibria (refer to the chaos model in ~\cite{Amaral2006}). However, in numerous applications, the difficulty arises when we lack prior information about the location of the subregions and the quantity of local modes to reach a particular approximation accuracy. 

The conceptualization of PWA approximation as an optimization problem becomes notably intricate when dealing with multi-dimensional nonlinear functions. Even for known, yet multi-dimensional models such as in resistor networks~\cite{Ohtsuki1977}, analytical solution of the PWA approximation problem may be elusive and the optimization problem should better be formulated for a set of points sampled from the domain~\cite{Wen2008}. This idea resembles PWA approximation approaches learned through experimental data~\cite{Bemporad2023}. A question arise about how to sample the points. A trivial solution is taking as many subregions as data points~\cite{Lauer2019}; however, this approach easily leads to overfitting. Therefore, for the optimization problem to become well-posed, the number of local modes is often fixed while minimizing approximation error~\cite{Lauer2019} or its expectation~\cite{Tang2021}. 

Various methods have been used to formulate the PWA approximation problem~\cite{Paoletti2007,Lauer2019,Moradvandi2023}. A common formulation is a bi-level optimization problem~\cite{Paoletti2019} that can be recast into a mixed-integer program~\cite{Bard2011}, or solved in a recursive manner~\cite{Mattsson2016,Mattsson2018,Breschi2016,Jin2021}. While recursive solutions are fast and can be used for online PWA approximation~\cite{Du2021}, they are often limited in handling multi-dimensional systems and are most effective when the form of the subregions is partially known and just needs to be refined~\cite{Kersting2019}. For instance, in~\cite{Oliveri2013} more vertices are iteratively added to the subregions for improved accuracy, but the solver needs to be properly initialized.

Bi-level optimization arises because PWA approximation essentially has two key aspects: establishing a partitioning strategy to divide the domain into subregions, and finding the local affine approximations. A popular partitioning strategy is clustering of the mesh points~\cite{Amaldi2016,Breschi2016,Du2020,Bako2014}, which can be sensitive to the mesh quality~\cite{Jin2021}. Despite the efforts to reduce the sensitivity to the cluster boundary and outliers~\cite{Khanmirza2016}, the performance of clustering-based partitioning degrades for multi-dimensional nonlinear functions. Some formulations use a specific shape for the partition, e.g.\ using hyper-rectangular subregions for digital systems~\cite{Comaschi2012}, using the function gradient~\cite{Azuma2010}, which is only applicable for uni-dimensional domains, or simplical representation~\cite{Bemporad2011}, which is applicable for low-dimensional domains~\cite{Paoletti2007}. Conversely, the hinging hyperplanes formalism, where the function is defined as a sum of hinging functions, e.g.\ $\min$ and $\max$, of parameterized hyperplanes, have proven to be an efficient and tractable formulation for partitioning of multi-dimensional domains~\cite{Roll2004}.

If the approximation problem is solved offline, the main goal is to converge to the most accurate PWA approximation with minimal complexity. This essentially requires a partitioning strategy that is flexible enough to divide the domain in various ways and consequently, allow for finding a PWA approximation of a desired accuracy via defining the lowest possible number of subregions. Additionally, \blue{in many control or optimization applications, discontinuity in the function may lead to undesirable behavior and/or numerical instabilities. Therefore, it is crucial that the resulting PWA function is continuous across its domain to ensure that no abrupt changes or discontinuities occur at the boundaries between adjacent affine subdomains/local modes. Therefore,} an appropriate methodology should be able to yield a continuous PWA approximation, which was handled by using min and max operators in~\cite{Xu2016}. However, using max and min operators does not allow for a discontinuous form. In this sense, using the continuity constraint from~\cite{Ohtsuki1977} is more suitable for more flexibility.

In this paper, we present a novel approach for PWA approximation of multi-dimensional nonlinear systems using cut-based partitioning of the domain via a hinging hyperplane formalism \blue{that allows for iterative increase of the PWA complexity level}. The complexity level of the PWA form can be iteratively increased if the intended level of complexity is either unknown or not restricted. In our earlier work~\cite{IFAC2023}, we have tackled the PWA approximation of multi-dimensional nonlinear systems with no prior knowledge. In this paper, we generalize the cutting definitions and the formulation of the optimization problem, enhancing the flexibility of the partitioning strategy. These extensions enable evaluating cuts that were not possible with our prior approach, and allow for a tractable implementation of continuity constraints in the approximation problem. In summary, our novel PWA approximation approach improves upon the state-of-the-art methods by (1) flexible definition of subregions using a generalized hinging hyperplane formalism to allow for finding PWA approximations with fewer number of subregions for a given approximation error tolerance, and (2) tractable bi-level formulation of the optimization problem to facilitate modifications based on applications e.g.\ imposing the continuity constraint on the PWA approximation. The efficacy of these extensions will be shown via illustrative examples. Further, we compare our method to the approach in~\cite{Jin2021} where a recursive solution of the PWA approximation problem helps reducing the complexity and we show that our optimization formulation and cut-based partitioning allows for convergence to the same accuracy level with fewer number of subregions. \blue{A modular version of our code is accessible through our GitHub repository\footnote{\texttt{\blue{https://github.com/leilagharavi/ cut-based-pwa-approximation}}}.}

The structure of the paper is as follows: Section~\ref{sec:prob} covers the main definitions and the formulation of the approximation problem. The novel cut-based partitioning strategy is explained in Section~\ref{sec:part} using examples for clarity, and Section~\ref{sec:pwa} presents the resulting optimization problem and the solution procedure. Case studies and comparisons are described in Section~\ref{sec:case}, while Section~\ref{sec:conc} concludes this paper.

\section{Problem Formulation}\label{sec:prob}

Let us consider a nonlinear system with dynamics
\[ \dot{s} = F (s,u),\]
where $s \in \mathbb{R}^n$ and $u \in \mathbb{R}^m$ respectively represent the state and input vectors and $F: \mathbb{R}^{n+m} \to \mathbb{R}^n$ is the nonlinear function to be approximated. Without loss of generality, the augmented state vector $x = [ s^T \; u^T]^T$ is used to define $F(x) \coloneqq F(s,u)$ since the approximated function will be selected to be affine in both the state and the input. The augmented domain is assumed to be bounded and is defined as $\dom(F) = \Dcal \subset \mathbb{R}^{n+m}$. For brevity of the expressions, hereafter we will use \mbox{$d=n+m$} as the dimension of the domain.

\begin{defn}[Domain $\Dcal$] The domain $\Dcal \subset \mathbb{R}^d$ is defined by the scalar boundary function $g: \mathbb{R}^d \to \mathbb{R}$ as
	\[\Dcal \coloneqq \{ x \in \mathbb{R}^d \; | \; 0 \leqslant g(x) \leqslant 1\}.\]
\end{defn} 

\begin{rem}
	We use the normalized form $0 \leqslant g \leqslant 1$ instead of $g \geqslant 0$ to avoid numerical issues in solving the approximation optimization problem.
\end{rem}

The proposed approach approximates the nonlinear function $F$ by a PWA function $f$ defined as
\begin{equation} \label{eq:pwadef}
    x \in \Ccal_p \implies f(x) = f_p (x), \quad \quad  f_p (x) = J_p x + K_p,
\end{equation}
with $p \in \{1, 2, \dots, P\}$, where the matrices $J_p \in \mathbb{R}^{n \times d}$ and the vectors $K_p \in \mathbb{R}^{n}$ describe in total $P$ local affine modes, each defined on a polytopic subregion $\Ccal_p \subset {R}^d$ such that the polytopes form a partition of $\Dcal$, i.e.\ the subregions are nonempty,
\begin{subequations}\label{eq:regs}
    \begin{equation}\label{eq:reg1}
        \Ccal_p \neq \emptyset, \; \forall p \in \{1, \dots, P\}
    \end{equation}  
    they are non-overlapping, 
    \begin{equation}\label{eq:reg2}
        \text{int}\left(\Ccal_p\right) \cap \text{int}\left(\Ccal_q\right) = \emptyset, \; \forall p, q \in \{1, \dots, P\}, \; p \neq q
    \end{equation}
    and their union covers the entire domain, 
    \begin{equation}\label{eq:reg3}
        \bigcup_{p=1}^{P} \Ccal_p = \Dcal,
    \end{equation}
\end{subequations}
with $\mathrm{int}(\Ccal_{p})$ denoting the interior of region $\Ccal_{p}$. 
\begin{defn}[Region set $\Rcal$] 
	The region set $\Rcal$ is the ordered set collecting the partition (i.e.\ the subregions) as
	\[\Rcal \coloneqq \{\Ccal_1, \Ccal_2, \dots, \Ccal_P\}.\] 
\end{defn}
For a fixed $P$, the region set $\Rcal$ and the corresponding local affine approximations $f_p$ are obtained simultaneously via solving the optimization problem
\begin{align} \label{eq:optprob}
	&\min_{\Jcal, \ \Kcal, \ \Rcal} & \; & \int\limits_{\Dcal} \dfrac{\left\Vert F(x) -f(x) \right\Vert_2^2}{\left\Vert F(x) \right\Vert_2^2 + 1}\; d x, \\ &\text{subject to}  &  & (\ref{eq:pwadef}) - (\ref{eq:regs}),
\end{align}
to minimize the squared approximation error where $\Jcal$ and $\Kcal$ represent the ordered sets containing $J_p$ and $K_p$, respectively. The term $\left\Vert F(x) \right\Vert^2_2$ in the denominator is introduced such that the cost values represent the relative error and the added 1 prevents division by very small values where $\Vert F(x) \Vert_2 \approx 0$.

\section{Parametric Region Definition}\label{sec:part}

We partition $\Dcal$ into $P$ subregions by cutting it using $(d-1)$--dimensional hyperplanes.
\begin{defn}[Cutting hyperplane $H_i$] 
	The $i$-th cutting hyperplane $H_i$ is an affine subspace of $\mathbb{R}^d$ defined as
	\[H_i \coloneqq \{ x \; | \; h_i^T x - 1 = 0\},\]
	for $i \in \{1,\dots,\nc\}$ where $n_{\mathrm{c}}$ represents the number of cuts.
\end{defn}
\begin{defn}[Cut arrangement $\Hcal$] 
	The cut arrangement $\Hcal$ is the arrangement of $\nc$ cutting hyperplanes defined by the set
	\[\Hcal = \{H_1, H_2, \dots, H_{\nc}\}.\]
\end{defn}

\begin{rem}
	In principle, the number of subregions generated by cutting $\mathbb{R}^d$ via the arrangement $\Hcal$ can be calculated using Zaslavsky's theorem~\cite{Zaslavsky1975} provided that all the possible 0- to $(d-1)$-dimensional intersections of the hyperplanes in $\Hcal$ are obtained. As a more computationally-efficient approach, here we fix the number of cutting hyperplanes and numerically obtain the region set $\Rcal$ within $\Dcal$ by investigating the existence of the possible subregions created by $\Hcal$ without counting the 0- to $(d-1)$-dimensional intersections. As a result,  $P$  is not fixed a priori.
\end{rem}

\begin{figure}[htbp]
	\begin{center}
		\begin{subfigure}{0.3\textwidth}\centering
			\includegraphics[width=0.65\textwidth]{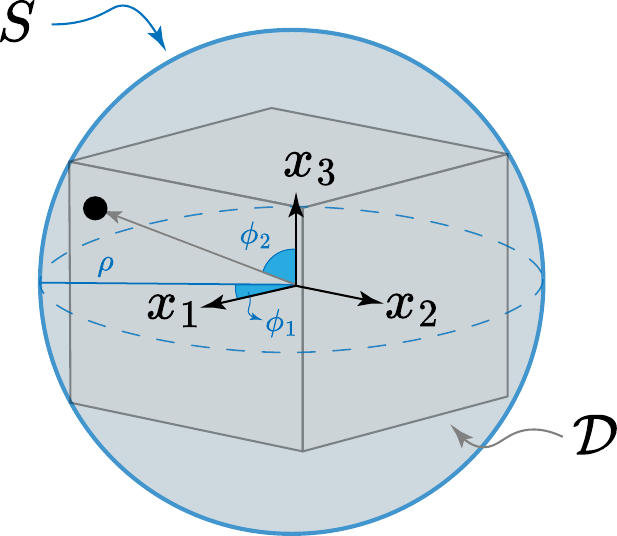}
			\subcaption{Definition of points on $S$}\label{fig:def1}
		\end{subfigure}
		\par\bigskip
		\begin{subfigure}{0.3\textwidth}
			\includegraphics[width=\textwidth]{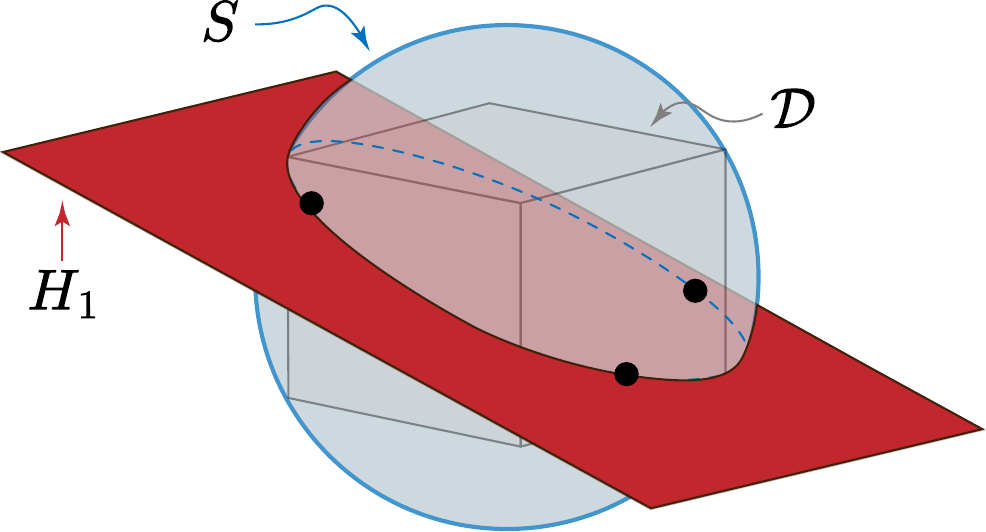}
			\subcaption{Definition of $H_1$ using $d$ points on $S$}\label{fig:def2}
		\end{subfigure}
		\par\bigskip
		\begin{subfigure}{0.3\textwidth}
			\includegraphics[width=\textwidth]{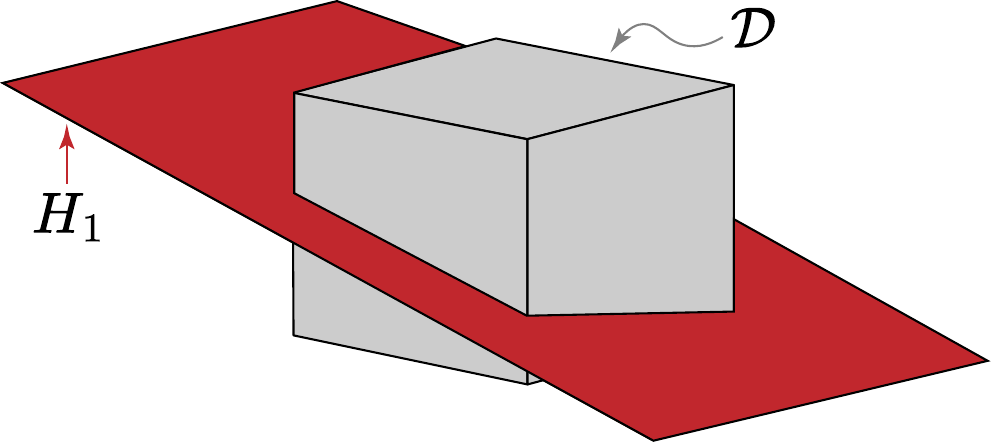}
			\subcaption{Cutting $\mathcal{D}$ using $H_1$}\label{fig:def3}
		\end{subfigure}
	\end{center}
	\caption{A schematic view cutting the domain using a hypersphere.}
	\label{fig:defs}
\end{figure}

To define each cutting hyperplane, we generate $d$ points in $\mathbb{R}^d$ and find the hyperplane passing through them as shown in Fig.~\ref{fig:defs}. These points are defined on the surface of a enclosing hypersphere $S$ in $\Dcal$ to ensure they are linearly-independent.
\begin{defn}[Enclosing hypersphere $S$] The enclosing hypersphere $S$ is the smallest $d$--dimensional hypersphere enclosing $\Dcal$ defined by
	\[S \coloneqq \{ x \; | \; \Vert x \Vert_2^2 - \rho^2 = 0\},\]
	where the constant $\rho$ is the radius of the enclosing hypersphere. 
\end{defn}
\begin{rem}
	Without loss of generality, one can always define a coordinate shift for the domain so that the center of the hypersphere is located at the origin. . This allows to simplify the expressions and mathematical manipulations.
\end{rem}

To locate the points on $S$, we use spherical coordinates $(r = \rho,\phi_1, \phi_2, \dots, \phi_{d-1})$ to parameterize the locations using spatial angles
\[\phi_1, \phi_2, \dots, \phi_{d-2} \in [0,\pi], \quad \phi_{d-1} \in [0,2\pi].\]
Figure~\ref{fig:def1} shows this concept for $d=3$. The Cartesian coordinates of the $k$-th point $x_k = [x_{1,k}, \ldots, x_{d,k}]^T$ can then be obtained by~\cite{Blumenson1960}
\begin{subequations}\label{eq:xks}
	\begin{equation}
		x_{j,k} = \rho \cos{\phi_{j,k}} \prod_{\nu=1}^{j-1} \sin{\phi_{\nu,k}}, \quad \forall j \in \{1, \dots, d-1\},
	\end{equation}
	\begin{equation}
		x_{d,k} = \rho \prod_{\nu=1}^{d-1} \sin{\phi_{\nu,k}},
	\end{equation}
\end{subequations}
where $x_{j,k}$ is the $j^{\rm{th}}$ component of the $k^{\rm{th}}$ point. Then, to find the hyperplane $H_i$ passing by $d$ points, we need to solve
\begin{equation}
	X_i h_i = \bold{1}_{d\times 1} \implies h_i = X_i^{-1} \bold{1}_{d\times 1},
	\label{eq:findH}
\end{equation}
where
\[X_i = \begin{bmatrix}
	x_{1,1} & x_{1,2} & \dots & x_{1,d}\\
	x_{2,1} & x_{2,2} & \dots & x_{2,d}\\	
	\vdots & & & \\
	x_{d,1} & x_{d,2} & \dots & x_{d,d}\\	
\end{bmatrix}.\]
Figure~\ref{fig:def2} illustrates the generation of one hyperplane $H_1$ for $d=3$ and how the domain is cut into two partitions in Fig.~\ref{fig:def3}. For more cuts, we then need to proceed analogously for all $\nc$ cuts, obtaining $\Hcal$ as
\begin{align}\label{eq:hcaldef}
	\Hcal = \{\bold{1}_{1\times d} \; X_i^{-T} x = 1 \}, && \forall i \in \{1, \dots, \nc\}.
\end{align}

Since each hyperplane divides the domain into two half-spaces, we define the map $\sigma$ from the domain $\Dcal$ to the $\nc \times 1$ Boolean vector $\sigma$ as 
\begin{align}
	\sigma_i (x) = \begin{cases}
		0 & \qquad \mathrm{if} \quad h_i^T x < 1\\ 1 & \qquad \mathrm{if} \quad  h_i^T \geqslant 1
	\end{cases}, \qquad \forall i \in \{1, \dots, \nc\},
	\label{eq:sigmaforH}
\end{align}
to indicate which side of the hyperplane $H_i$ the point $x$ lies on. Since the subregions are also located on one side of each hyperplane, there exist at most $2^{\nc}$ possible partitions that can be stored in an $\nc \times 2^{\nc}$ matrix. However, to avoid unnecessary usage of memory, we suggest generating $\sigma$ vectors by investigating the binary vectors corresponding to integer numbers from $0$ to $2^{\nc}-1$ without storing all of them in a very large matrix. We use the prune-and-search paradigm~\cite{Edelsbrunner1987} by solving $2^{\nc}$ linear programs to check the feasibility of each combination 
\begin{align} 
	&\min_{x} & \; & 1, \label{eq:partfeasj}\\ 
	&\text{subject to}  &  & x \in \Dcal,\label{eq:partfeasc1}\\
	&  &  & h_i^T x < 1 \quad \text{if } \sigma_i = 0, \; i \in \{1, \dots, \nc\},\label{eq:partfeasc0}\\
	&  &  & h_i^T x \geqslant 1 \quad \text{if } \sigma_i = 1, \; i \in \{1, \dots, \nc\},\label{eq:partfeasc2}
\end{align}
where $\sigma \in \mathbb{R}^{\nc}$ is the binary representation of the integer $j \in \{0, \dots, 2^{\nc}-1\}$ in each linear program. The $\sigma$ vectors corresponding to feasible problems are then stored in the feasibility matrix\footnote{In the combination geometry literature, a similar concept is used but in a set of tuples called the oriented metroid. For more details see~\cite{Orlik2013}.}$\Sigma \in \mathbb{R}^{\nc \times P }$. This procedure is implemented by Algorithm~\ref{alg:feas}.
\begin{algorithm}
	\caption{Find feasibility matrix of a cut arrangement as $\Sigma = \texttt{chambers} (\Hcal,\Dcal)$}\label{alg:feas}
	\begin{algorithmic}
		\Require $\Hcal, \Dcal$
		\State $\nc \gets \vert \Hcal \vert$ \Comment{\emph{$\vert . \vert$ denotes cardinality}}
		\State $\sigma \gets 0_{\nc \times 1}$
		\State $\Sigma \gets 0_{\nc \times 0}$ \Comment{\emph{$0_{\nc \times 0}$ is an empty matrix}}
		\For{$l \in \{0, 1, \dots, 2^{\nc}-1\}$}
		\State $\sigma \xleftarrow{\text{ binary vector }} l$
		\If{(\ref{eq:partfeasj})-(\ref{eq:partfeasc2}) feasible for $(\sigma, \Hcal)$}
		\State $\Sigma \gets \begin{bmatrix} \Sigma & \sigma
		\end{bmatrix}$ 
		\EndIf
		\EndFor
		\State \Return $\Sigma$
	\end{algorithmic}
\end{algorithm}
\begin{exmp}
	Consider the 3-dimensional hypercube domain $\Dcal$ shown in Fig.~\ref{fig:cutinout} and two cut arrangements $\Hcal_a$ and $\Hcal_b$ respectively shown in Figures~\ref{fig:cutout} and~\ref{fig:cutin}. We have 
	\[\nc = 2 \quad \implies l \in \{0,1,2,3\}\xrightarrow{\mathrm{binary}}  \{00,01,10,11\}.\] 
	Problem (\ref{eq:partfeasj})-(\ref{eq:partfeasc2}) is not feasible for $\sigma=01$ for $\Hcal_a$ as there is no region within $\Dcal$ lying above $H_1$ and below $H_2$. However, (\ref{eq:partfeasj})-(\ref{eq:partfeasc2}) is feasible for $\sigma \in \{00,10,11\}$. Therefore, the feasibility matrices for $\Hcal_a$ and $\Hcal_b$ are
	\[\Sigma_a = \begin{bmatrix} 0 & 1 & 1 \\ 0 & 0 & 1 \end{bmatrix}, \qquad 
	\Sigma_b = \begin{bmatrix} 0 & 0 & 1 & 1 \\ 0 & 1 & 0 & 1 \end{bmatrix}.\]
	\begin{figure}[htbp]
		\begin{center}
			\begin{subfigure}{0.23\textwidth}
				\includegraphics[width=\textwidth]{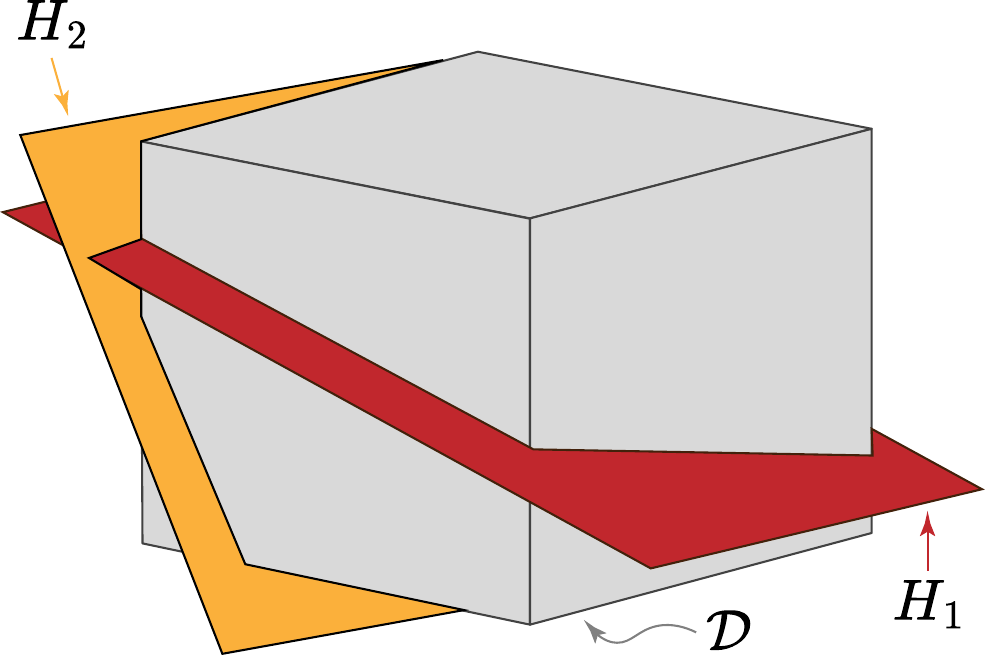}
				\subcaption{$\Hcal_a$}\label{fig:cutout}
			\end{subfigure}
			\begin{subfigure}{0.23\textwidth}
				\includegraphics[width=\textwidth]{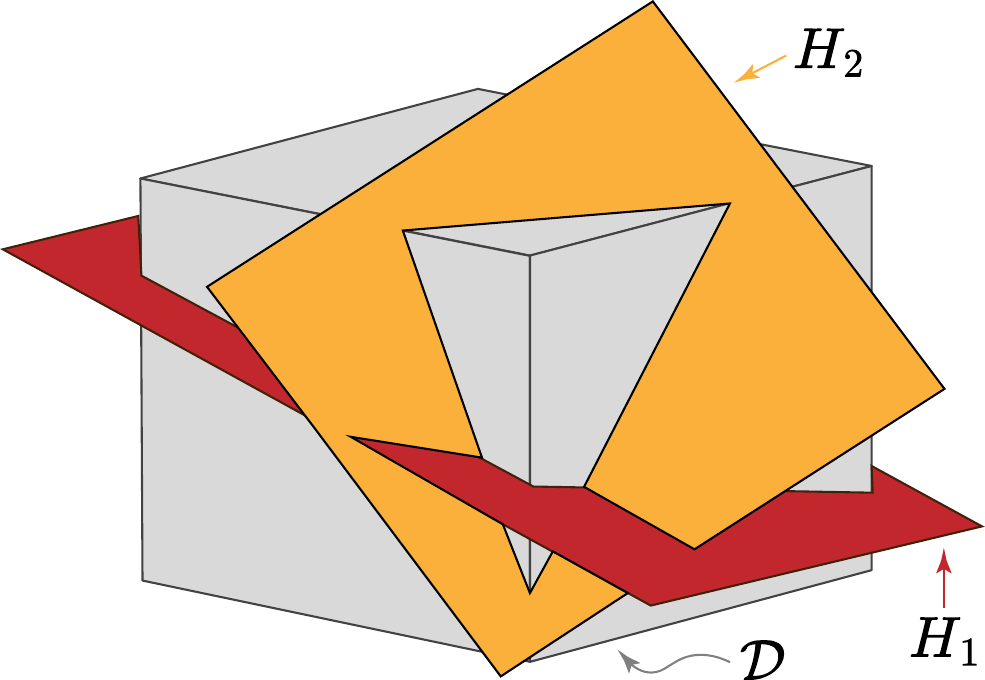}
				\subcaption{$\Hcal_b$}\label{fig:cutin}
			\end{subfigure}
		\end{center}
		\caption{A schematic view of two cut arrangements in Example~1.}
		\label{fig:cutinout}
	\end{figure}
\end{exmp}

Next, we aim at finding the neighboring subregions to identify the hyperplane they share as their boundary. We define the adjacency matrix $A$ to store this information.
\begin{defn}[Adjacency Matrix $A$] 
	The adjacency matrix $A \in \mathbb{R}^{P \times P}$ represents the neighboring subregions within the region set $\Rcal$ as follows:
	\[A_{p,q} = \begin{cases}i & \qquad \mathrm{if} \quad \Ccal_p \cap \Ccal_q = H_i \\ 0 & \qquad  \mathrm{otherwise}\end{cases}.\]
	The adjacency matrix is symmetric by definition.
\end{defn}
Note that the adjacent subregions share their $\sigma$ vector, except for only one element, which is the element corresponding to their boundary hyperplane. Thus, the adjacency matrix is constructed by investigating the columns in $\Sigma$. Since each column in $\Sigma$ is a binary vector, two columns differ only in one element if and only if their subtraction contains only one $\pm 1$ element, i.e.\ 
\[\Vert \Sigma_{.,p} - \Sigma_{.,q} \Vert_1 = 1.\]
Each column of $\Sigma$ represents a subregion and how it relates to the hyperplanes in $\Hcal$. However, defining each region is done only by evaluating its boundaries and not all the cutting hyperplanes to avoid redundancy. Since each column (or row) of $A$ corresponds to one of the elements in $\Rcal$, the subregion $p$ can now be formulated as follows:
\begin{multline}
	\Ccal_p = \Big\{ x \in \Dcal, \;  \forall i>0, \; A_{p,.} = i  \; \Big| \\ (-1)^{\Sigma_{i,p}} \; h_i^T x \leqslant (-1)^{\Sigma_{i,p}}\Big\}.\label{eq:findccal}
\end{multline}
The subregions are then stored in the region set $\Rcal$. Algorithm~\ref{alg:reg} describes the procedure of obtaining $\Rcal$ from $\Hcal$ and $\Sigma$ using the adjacency matrix $A$.
\begin{algorithm}
	\caption{Find subregions of a cut arrangement as $(\Rcal,A,P) = \texttt{regions} (\Hcal, \Sigma)$ }\label{alg:reg}
	\begin{algorithmic}
		\Require $\Hcal, \Sigma$
		\State $\nc \gets \vert \Hcal \vert$ \Comment{\emph{$\vert . \vert \coloneqq$ cardinality}}
		\State $P \gets \text{number of columns in } \Sigma$
		\State $A \gets 0_{P \times P}$
		\State $\Rcal \gets \emptyset$
		\For{$p \in \{1, \dots, P\}$}
		\For{$q \in \{p+1, \dots, P\}$}  
		\State $\delta \gets \Sigma_{.,p} - \Sigma_{.,q}$ 
		\If{$\Vert \delta \Vert_1 = 1$}
		\State $A_{p,q} \gets$ index of nonzero component in $\delta$
		\EndIf
		\EndFor
		\EndFor
		\State $A \gets A + A^T$ \Comment{\emph{upper-triangular to symmetric}}
		\For{$p \in \{1, \dots, P\}$}
		\State $\Ccal_p \xleftarrow{\text{ solve (\ref{eq:findccal}) }} \Hcal , A$
		\State $\Rcal \gets \Rcal \cup \{\Ccal_p\}$
		\EndFor
		\State \Return $\Rcal, A, P$
	\end{algorithmic}
\end{algorithm}
	\begin{figure}[htbp]
	\begin{center}
		\begin{subfigure}[t]{0.23\textwidth}
			\includegraphics[width=\textwidth]{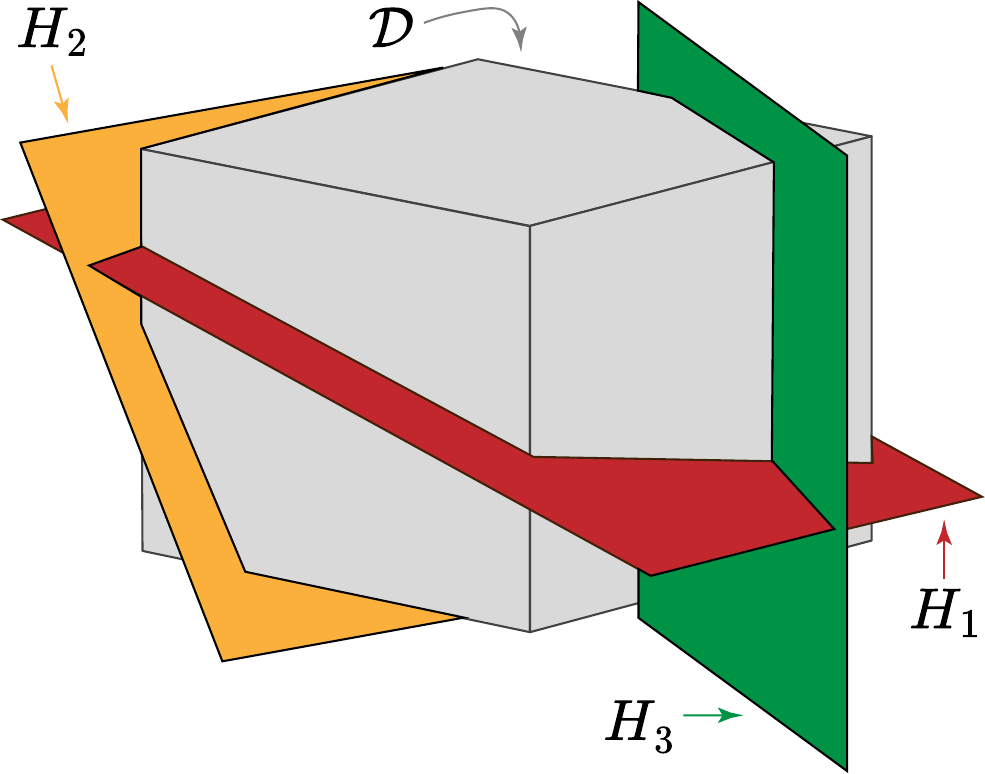}
			\subcaption{Cutting the domain}\label{fig:adj}
		\end{subfigure}
		\begin{subfigure}[t]{0.23\textwidth}
			\centering
			\includegraphics[width=0.8\textwidth]{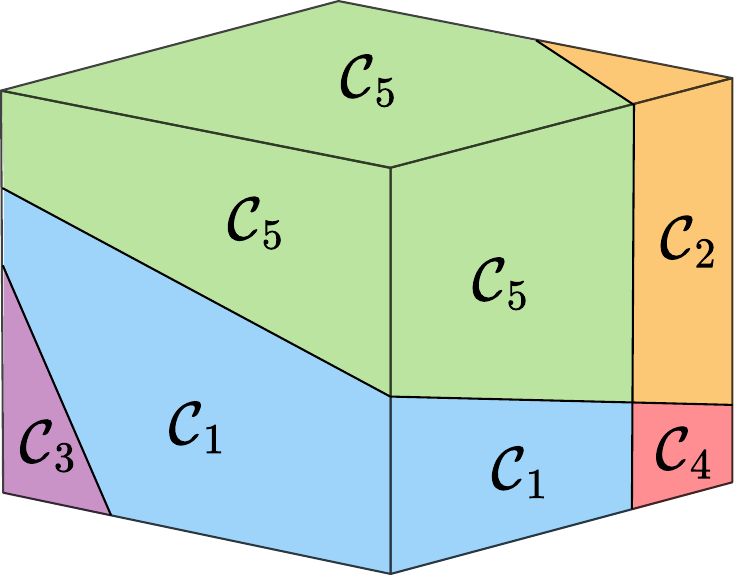}
			\subcaption{Resulting subregions}\label{fig:reg}
		\end{subfigure}
	\end{center}
	\caption{Illustration of the cut arrangement and the resulting subregions in Example 2.}
	\label{fig:exmpreg}
\end{figure}
\begin{exmp}
	Consider the 3-dimensional domain shown in Fig.~\ref{fig:exmpreg} as
	\[\Dcal \coloneqq \{\; x \in \mathbb{R}^3 \; \; \vert \; \; x_j \in [-2,2] \; , \; j \in \{1,2,3\} \},\]
	with the cut arrangement $\Hcal = \{H_1, H_2, H_3\}$ shown in Fig.~\ref{fig:adj} where 
	\begin{align*}
		&H_1 \coloneqq \{\; x \in \mathbb{R}^3 \; \; \vert \; \; -x_1 + 2x_2 + 5x_3 = 1\},\\
		&H_2 \coloneqq \{\; x \in \mathbb{R}^3 \; \; \vert \; \; 0.1 x_1 - 0.5 x_2 - 0.2 x_3 = 1\},\\
		&H_3 \coloneqq \{\; x \in \mathbb{R}^3 \; \; \vert \; \; - x_1 + x_2 = 1\}.
	\end{align*}
	The feasibility and adjacency matrices for $\Hcal$ are
	\[\Sigma = \begin{bmatrix} 0 & 0 & 0 & 1 & 1\\0 & 0 & 1 & 0 & 0\\0 & 1 & 0 & 0 & 1 \end{bmatrix}, \qquad A = \begin{bmatrix} 
		0 & 3 & 2 & 1 & 0\\
	 	3 & 0 & 0 & 0 & 1\\
  		2 & 0 & 0 & 0 & 0\\
 		1 & 0 & 0 & 0 & 3\\
 		0 & 1 & 0 & 3 & 0\\ \end{bmatrix},\]
 	which gives the subregions as depicted in Fig.~\ref{fig:reg} expressed by
	\begin{align*}
		&\Ccal_1 \coloneqq \{\; x \in \Dcal \; \; \vert \; \; \begin{bmatrix}
			-1 & 2 & 5 \\ 0.1 & -0.5 & -0.2 \\ -1 & 1 & 0 
		\end{bmatrix} x \leqslant 1\},\\
		&\Ccal_2 \coloneqq \{\; x \in \Dcal \; \; \vert \; \; \begin{bmatrix}
		-1 & 2 & 5 \\ 1 & -1 & 0 
		\end{bmatrix} x \leqslant 1\},\\
		&\Ccal_3 \coloneqq \{\; x \in \Dcal \; \; \vert \; \; \begin{bmatrix}
		-0.1 & 0.5 & 0.2 
		\end{bmatrix} x \leqslant 1\},\\
		&\Ccal_4 \coloneqq \{\; x \in \Dcal \; \; \vert \; \; \begin{bmatrix}
			1 & -2 & -5 \\ -1 & 1 & 0 
		\end{bmatrix} x \leqslant 1\},\\
		&\Ccal_5 \coloneqq \{\; x \in \Dcal \; \; \vert \; \; \begin{bmatrix}
			1 & -2 & -5 \\ 1 & -1 & 0 
		\end{bmatrix} x \leqslant 1\},\\
	\end{align*}
\end{exmp}

\section{Cut-Based PWA Approximation}\label{sec:pwa}

The PWA approximation of a nonlinear function is obtained by solving (\ref{eq:optprob}) by optimizing the partitioning strategy via $\Rcal$ and the local affine approximations $\Jcal$ and $\Kcal$, while satisfying (\ref{eq:regs}). Using the cut-based parametric region definition through Algorithms~\ref{alg:feas} and~\ref{alg:reg}, constraints (\ref{eq:regs}) are automatically satisfied since we cut the domain with hyperplanes which gives $\Rcal$ as a set of non-overlapping partitions that cover the whole $\Dcal$.

If the aim is to give a continuous PWA approximation of the system, the following constraint should be imposed on the dynamic modes corresponding to neighboring subregions, which is derived based on~\cite{Fujisawa1972}: 
\begin{equation}
	\begin{aligned}
		&\exists c \in \mathbb{R}^d \quad \text{s.t.} \quad  J_p - J_q = c \; h_i^T, \\ &\qquad\qquad \forall p, q \in \{1, \dots, P\}, \; H_i = \Ccal_p \cap \Ccal_q.
	\end{aligned}\label{eq:fujicont}
\end{equation}
\begin{cor}
	The difference of $J_p$ and $J_q$ in (\ref{eq:fujicont}) is of rank one~\cite{Fujisawa1972}. As a result, the continuity can be imposed on the PWA approximation by considering rank of the Jacobian matrices in neighboring modes. 
\end{cor}

Even with a fixed number of cutting hyperplanes, the number of subregions (and consequently the dimensions of the decision space) differs for different $\Hcal$ arrangements with the same $\nc$. Due to its varying-dimensional nature, we formulate the PWA approximation problem as a bi-level optimization problem. At the lower level, we find the minimum approximation error for the region set $\Rcal$ as
\begin{align}
	\Gamma^\ast (\Rcal) = \; & \min_{\Jcal,\Kcal} && \sum_{p=1}^{P}\int\limits_{\Ccal_p} \dfrac{\left\Vert F(x) - J_p x - K_p \right\Vert_2^2}{\left\Vert F(x) \right\Vert_2^2 + 1}\; d x, \label{eq:lowlevelj} \\
	& \text{ s.t.} &&  \mathrm{rank} \left(J_p - J_q \right) = 1, \label{eq:lowlevelc}\\
	& && \hspace{20pt} \forall p, q \in \{1, \dots, P\}, \; A_{p,q} \neq 0, \notag
\end{align}
where (\ref{eq:lowlevelc}) is the continuity constraint and can be disregarded in case a continuous PWA form is not required. Problem (\ref{eq:lowlevelj})-(\ref{eq:lowlevelc}) is a nonlinear least-squares optimization problem that can be solved using e.g.\ gradient-based methods with multiple starting points.

At the higher level, we solve the following optimization problem:
\begin{align}
	& \min_{\phi} && \Gamma^\ast (\Rcal) + \lambda P \label{eq:highlevelj}\\
	& \text{ s.t.} && \Hcal = \{\bold{1}_{1\times d} \; X_i^{-T} x = 1 \}, \label{eq:highlevelc1} \\
	& && \hspace{70pt} \forall i \in \{1, \dots, \nc\}, \notag\\
	& && \Sigma = \texttt{chambers} (\Hcal,\Dcal), \label{eq:highlevelc2} \\
	& && (\Rcal, A, P) = \texttt{regions} (\Hcal, \Sigma), \label{eq:highlevelc3}\\
	& && \phi_{i,j} \in [0,\pi], \quad \phi_{i,d-1} \in [0,2\pi], \label{eq:highlevelb}\\
	& && \hspace{70pt} \forall i \in \{1, \dots, \nc\}, \notag\\	
	& && \hspace{70pt} \forall j \in \{1, \dots, d-2\}, \notag
\end{align}
with $\phi$ collecting $\phi_{i,j}$, $\Gamma^\ast$ being the approximation error obtained in the lower level and the \texttt{chambers}\footnote{The term ``chambers" is often used in the combinatorial geometry literature for the ``subregion" concept. Here, to distinguish between the functions in Algorithms~\ref{alg:feas} and \ref{alg:reg}, we use this term as a label for clarity.} and  \texttt{regions} functions correspond to the Algorithms~\ref{alg:feas} and \ref{alg:reg}, respectively. Here we penalize the number of subregions $P$ as well by a regularizing weight \mbox{$\lambda > 0$}. Similar to the lower-level problem, (\ref{eq:highlevelj})-(\ref{eq:highlevelb}) is a nonlinear optimization problem. However, being not smooth, we propose solving it using global optimization methods such as genetic algorithm.

Equations (\ref{eq:lowlevelj})-(\ref{eq:highlevelb}) are solved for a fixed $\nc$. In our approach, we start by $\nc = 1$ and in case the best solution does not reach a user-defined error tolerance, we increase $\nc$ and solve (\ref{eq:lowlevelj})-(\ref{eq:highlevelb}) again. Algorithm~\ref{alg:pwa} describes this iterative cut-based PWA optimization problem.
\begin{algorithm}
	\caption{Iterative cut-based PWA approximation}\label{alg:pwa}
	\begin{algorithmic}
		\Require $F,\Dcal$
		\State generate $S$
		\State $\text{cond} \gets \text{true}$
		\State $\nc \gets 0$
		\State $\text{iter} \gets$ 0		
		\While{cond} 	
		\State $\nc \gets  \nc + 1$	
		\State iter $\gets$ iter $+1$
		\State $(\Gamma^\ast,\Jcal^\ast, \Kcal^\ast, \Rcal^\ast) \gets$ solve (\ref{eq:lowlevelj})-(\ref{eq:highlevelb})				
		\If{$ \Gamma^\ast \leqslant \text{tol}_\text{err}$} 
		\State $\text{cond} \gets \text{false}$
		\State \Return $\Jcal^\ast, \Kcal^\ast, \Rcal^\ast$
		\ElsIf {$\text{iter} \geqslant \text{iter}_\text{max}$} 
		\State $\text{cond} \gets \text{false}$
		\State \Return $\text{print}(\text{`Exceeded } \text{iter}_\text{max} \text{'})$
		\EndIf
		\EndWhile
	\end{algorithmic}
\end{algorithm}

\blue{\begin{rem}
		The computational complexity of Algorithm~\ref{alg:feas} (in terms of computation time) increases exponentially with the number of cuts $\nc$ and quadratically with the dimension $d$, while for Algorithm~\ref{alg:reg} the complexity increase is respectively quadratic and linear. Therefore, the complexity of Algorithm~\ref{alg:pwa} is expected to increase exponentially with $\nc$ and polynomially with $d$.
	\end{rem}}

\section{Results and Discussion}\label{sec:case}

In this section, we analyze the performance of our proposed cut-based PWA approximation method in three steps: first, we use a case study as an illustrative example of a nonlinear function for a user-defined approximation error tolerance. Then, we compare our method with~\cite{IFAC2023} and~\cite{Jin2021} to show the flexibility of our approach, leading to lower approximation errors for a lower complexity of PWA the form. 

\subsection{Case Study}

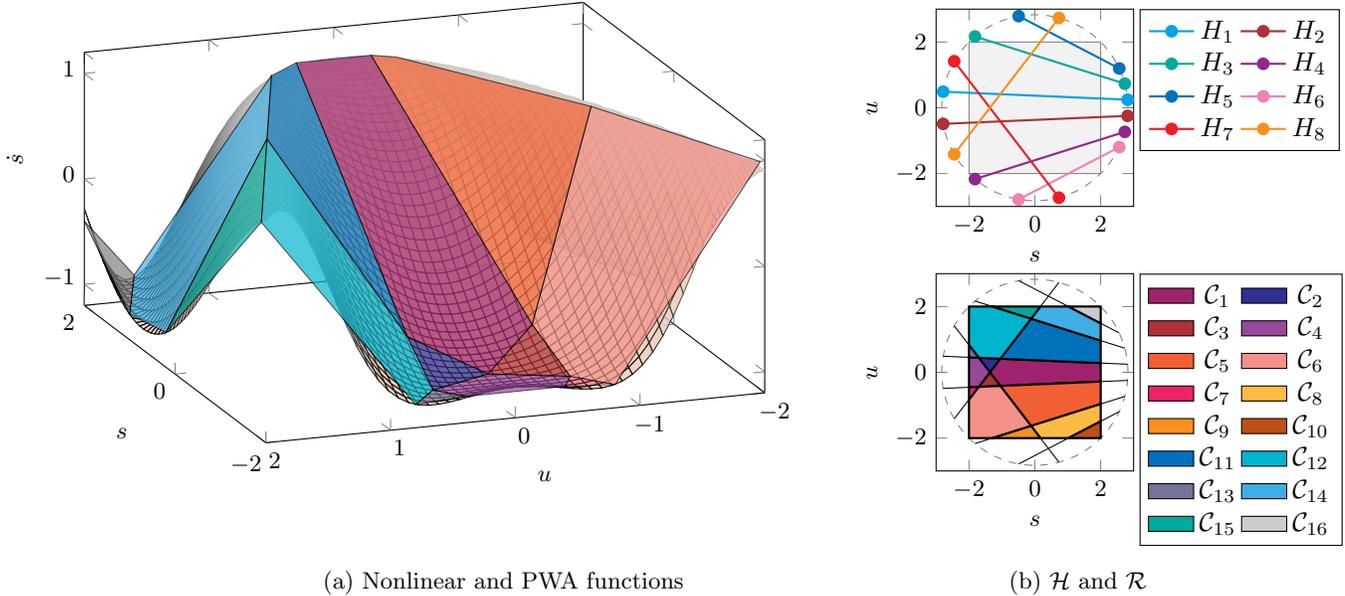
\begin{figure*}[hbt]
\begin{center}
\begin{subfigure}[b]{0.6\textwidth}
\begin{tikzpicture}
\hspace{-40pt}
\begin{axis}[height=0.7\textwidth,width=\textwidth,view={-110}{30},
	ylabel=$u$,xlabel=$s$,zlabel=$\dot{s}$,colormap/blackwhite]
	\addplot3[surf,draw=Tan,fill=Tan!50,fill opacity=0.8,samples=50,domain=-2:2,domain y=-2:2] {sin(deg(x+y^2)};	
	\coordinate (P1) at (2,-2,-0.3);
	\coordinate (P2) at (2,-1.5,-0.9);
	\coordinate (P3) at (1,-2,-0.96);
	\coordinate (P4) at (0.2,-2,-0.87);
	\coordinate (P5) at (-0.1,-1.3,0.98);
	\coordinate (P6) at (2,-0.5,0.86);
	\coordinate (P7) at (2,-0.3,0.9);
	\coordinate (P8) at (-1,-0.5,-0.7);
	\coordinate (P9) at (-1.7,-0.55,-0.97);
	\coordinate (P10) at (-2,-0.8,-0.98);
	\coordinate (P11) at (-2,-1.7,0.9);
	\coordinate (P12) at (-1.9,-2,0.96);
	\coordinate (P13) at (-1.4,0,-0.98);
	\coordinate (P14) at (-1,0.5,-0.7);
	\coordinate (P15) at (-1.7,0.55,-0.97);
	\coordinate (P16) at (-2,0.4,-0.95);
	\coordinate (P17) at (-2,-0.4,-0.95);
	\coordinate (P18) at (-2,0.8,-0.98);
	\coordinate (P19) at (-1.9,2,0.86);
	\coordinate (P20) at (-0.1,1.3,0.98);
	\coordinate (P21) at (0.2,2,-0.87);
	\coordinate (P22) at (1,2,-0.96);
	\coordinate (P23) at (2,1.6,-0.96);
	\coordinate (P24) at (2,0.5,0.8);
	\coordinate (P25) at (2,0.3,0.9);
	\coordinate (P26) at (2,2,-0.4);
	\coordinate (P27) at (-2,1.7,0.9);
    \draw[fill=RedViolet,thin,opacity=0.7] (P7)--(P8)--(P13)--(P14)--(P25)--(P7);    
	\draw[fill=Blue,thin,opacity=0.7] (P13)--(P14)--(P15)--(P13);    
	\draw[fill=Maroon,thin,opacity=0.7] (P8)--(P9)--(P13)--(P8);    
	\draw[fill=Purple,thin,opacity=0.7] (P17)--(P9)--(P13)--(P15)--(P16)--(P17); 
	\draw[fill=RedOrange,thin,opacity=0.7] (P7)--(P6)--(P5)--(P8)--(P7); 
	\draw[fill=Salmon,thin,opacity=0.7] (P10)--(P12)--(P5)--(P8)--(P9)--(P10); 
	\draw[fill=WildStrawberry,thin,opacity=0.4] (P9)--(P10)--(P17)--(P9);  
	\draw[fill=RoyalBlue,thin,opacity=0.7] (P25)--(P24)--(P20)--(P14)--(P25); 	
	\draw[fill=Turquoise,thin,opacity=0.7] (P18)--(P19)--(P20)--(P14)--(P15)--(P18);  	
	\draw[fill=CadetBlue,thin,opacity=0.7] (P15)--(P16)--(P18)--(P15);  
	\draw[fill=CornflowerBlue,thin,opacity=0.7] (P20)--(P21)--(P22)--(P23)--(P24)--(P20); 
	\draw[fill=Emerald,thin,opacity=0.7] (P20)--(P19)--(P21)--(P20); 
	\draw[fill=gray,thin,opacity=0.7] (P23)--(P22)--(P26)--(P23); 
\end{axis}
\end{tikzpicture}	
\vspace{10pt}
\subcaption{Nonlinear and PWA functions}\label{fig:casef}
\end{subfigure}
\begin{subfigure}[b]{0.25\textwidth}
\begin{tikzpicture}\hspace{-20pt}
\begin{axis}[xlabel={$s$},ylabel={$u$},xmin=-3,xmax=3,ymin=-3,ymax=3,
	height=0.95\textwidth,width=0.95\textwidth,legend columns=2,legend pos=outer north east]
	\draw[gray,thin,fill=gray!10] (-2,-2) rectangle (2,2);
	\draw[gray,thin,dashed] (0,0) circle [radius={2.*sqrt(2)}];
	\addplot+[thick,Cerulean,mark options={fill=Cerulean},solid,mark=*] coordinates {
		({2.83*cos(170)},{2.82*sin(170)}) ({2.83*cos(5)},{2.82*sin(5)})};
	\addplot+[thick,Maroon,mark options={fill=Maroon},solid,mark=*] coordinates {
		({2.83*cos(355)},{2.82*sin(355)}) ({2.83*cos(190)},{2.82*sin(190)})};
	\addplot+[thick,Emerald,mark options={fill=Emerald},solid,mark=*] coordinates {
		({2.83*cos(15)},{2.83*sin(15)}) ({2.83*cos(130)},{2.83*sin(130)})};
	\addplot+[thick,Plum,mark options={fill=Plum},solid,mark=*] coordinates {
		({2.83*cos(-15)},{2.83*sin(-15)}) ({2.83*cos(-130)},{2.83*sin(-130)})};
	\addplot+[thick,RoyalBlue,mark options={fill=RoyalBlue},solid,mark=*] coordinates {
		({2.83*cos(25)},{2.83*sin(25)}) ({2.83*cos(100)},{2.83*sin(100)})};
	\addplot+[thick,CarnationPink,mark options={fill=CarnationPink},solid,mark=*] coordinates {
		({2.83*cos(-25)},{2.83*sin(-25)}) ({2.83*cos(-100)},{2.83*sin(-100)})};
	\addplot+[thick,Red,mark options={fill=Red},solid,mark=*] coordinates {
		({2.83*cos(-75)},{2.83*sin(-75)}) ({2.83*cos(150)},{2.83*sin(150)})};
	\addplot+[thick,BurntOrange,mark options={fill=BurntOrange},solid,mark=*] coordinates {
		({2.83*cos(75)},{2.83*sin(75)}) ({2.83*cos(-150)},{2.83*sin(-150)})};
	\legend{$H_1$,$H_2$,$H_3$,$H_4$,$H_5$,$H_6$,$H_7$,$H_8$};
\end{axis}
\end{tikzpicture}
\begin{tikzpicture}\hspace{-20pt}
\begin{axis}[xlabel={$s$},ylabel={$u$},xmin=-3,xmax=3,ymin=-3,ymax=3,
	height=0.95\textwidth,width=0.95\textwidth,legend columns=2,legend pos=outer north east]
	\draw[gray,thin,dashed] (0,0) circle [radius={2.*sqrt(2)}];
	\addplot+[name path=uy,thin,black,no marks,solid,forget plot] coordinates {(-2,2) (2,2)};
	\addplot+[name path=ly,thin,black,no marks,solid,forget plot] coordinates {(-2,-2) (2,-2)};
	\addplot+[name path=ux,thin,black,no marks,solid,forget plot] coordinates {(2,-2) (2,2)};
	\addplot+[name path=lx,thin,black,no marks,solid,forget plot] coordinates {(-2,-2) (-2,2)};
	\addplot+[name path=H1,thin,black,no marks,solid,forget plot] coordinates {
		({2.83*cos(170)},{2.82*sin(170)}) ({2.83*cos(5)},{2.82*sin(5)})};
	\addplot+[name path=H2,thin,black,no marks,solid,forget plot] coordinates {
		({2.83*cos(355)},{2.82*sin(355)}) ({2.83*cos(190)},{2.82*sin(190)})};
	\addplot+[name path=H3,thin,black,no marks,solid,forget plot] coordinates {
		({2.83*cos(15)},{2.83*sin(15)}) ({2.83*cos(130)},{2.83*sin(130)})};
	\addplot+[name path=H4,thin,black,no marks,solid,forget plot] coordinates {
		({2.83*cos(-15)},{2.83*sin(-15)}) ({2.83*cos(-130)},{2.83*sin(-130)})};
	\addplot+[name path=H5,thin,black,no marks,solid,forget plot] coordinates {
		({2.83*cos(25)},{2.83*sin(25)}) ({2.83*cos(100)},{2.83*sin(100)})};
	\addplot+[name path=H6,thin,black,no marks,solid,forget plot] coordinates {
		({2.83*cos(-25)},{2.83*sin(-25)}) ({2.83*cos(-100)},{2.83*sin(-100)})};
	\addplot+[name path=H7,thin,black,no marks,solid,forget plot] coordinates {
		({2.83*cos(-75)},{2.83*sin(-75)}) ({2.83*cos(150)},{2.83*sin(150)})};
	\addplot+[name path=H8,thin,black,no marks,solid,forget plot] coordinates {
		({2.83*cos(75)},{2.83*sin(75)}) ({2.83*cos(-150)},{2.83*sin(-150)})};
	\path[name intersections={of=lx and H1,by=B10}];
	\path[name intersections={of=lx and H2,by=B20}];
	\path[name intersections={of=uy and H3,by=B30}];
	\path[name intersections={of=ly and H4,by=B40}];
	\path[name intersections={of=uy and H5,by=B50}];
	\path[name intersections={of=ly and H6,by=B60}];
	\path[name intersections={of=lx and H7,by=B70}];
	\path[name intersections={of=lx and H8,by=B80}];	
	\path[name intersections={of=ux and H1,by=B11}];
	\path[name intersections={of=ux and H2,by=B21}];
	\path[name intersections={of=ux and H3,by=B31}];
	\path[name intersections={of=ux and H4,by=B41}];
	\path[name intersections={of=ux and H5,by=B51}];
	\path[name intersections={of=ux and H6,by=B61}];
	\path[name intersections={of=ly and H7,by=B71}];
	\path[name intersections={of=uy and H8,by=B81}];
	\path[name intersections={of=lx and H1,by=B10}];
	\path[name intersections={of=lx and H2,by=B20}];
	\path[name intersections={of=H1 and H7,by=I17}];
	\path[name intersections={of=H2 and H8,by=I28}];
	\path[name intersections={of=H2 and H7,by=I27}];
	\path[name intersections={of=H1 and H8,by=I18}];
	\path[name intersections={of=H7 and H8,by=I78}];
	\path[name intersections={of=H4 and H7,by=I47}];
	\path[name intersections={of=H3 and H8,by=I38}];
	\filldraw[thick,fill=RedViolet] 
	(B11) -- (B21) -- (I27) -- (I78) -- (I18) -- cycle;
	\addlegendimage{area legend,fill=RedViolet};
	\filldraw[thick,fill=Blue] 
	(I18) -- (I78) -- (I17) -- cycle;
	\addlegendimage{area legend,fill=Blue};
	\filldraw[thick,fill=Maroon] 
	(I28) -- (I78) -- (I27) -- cycle;
	\addlegendimage{area legend,fill=Maroon};	
	\filldraw[thick,fill=Purple] 
	(B10) -- (B20) -- (I28) -- (I78) -- (I17) -- cycle; 
	\addlegendimage{area legend,fill=Purple};
	\filldraw[thick,fill=RedOrange] 
	(B21) -- (B41) -- (I47) -- (I27) -- cycle;	
	\addlegendimage{area legend,fill=RedOrange};
	\filldraw[thick,fill=Salmon] 
	(B40) -- (I47) -- (I27) -- (I28) -- (B80) -- (-2,-2) -- cycle;
	\addlegendimage{area legend,fill=Salmon};			
	\filldraw[thick,fill=WildStrawberry] 
	(B80) -- (B20) -- (I28) -- cycle;
	\addlegendimage{area legend,fill=WildStrawberry};
	\filldraw[thick,fill=gray,fill=Dandelion] 
	(B61) -- (B41) -- (I47) -- (B71) -- (B60) -- cycle;
	\addlegendimage{area legend,fill=Dandelion};
	\filldraw[thick,fill=BurntOrange] 
	(B40) -- (B71) -- (I47) -- cycle;
	\addlegendimage{area legend,fill=BurntOrange};
	\filldraw[thick,fill=Bittersweet] 
	(B60) -- (B61) -- (2,-2) -- cycle;
	\addlegendimage{area legend,fill=Bittersweet};
	\filldraw[thick,fill=RoyalBlue] 
	(B11) -- (B31) -- (I38) -- (I18) -- cycle;
	\addlegendimage{area legend,fill=RoyalBlue};
	\filldraw[thick,fill=Turquoise] 
	(B30) -- (I38) -- (I18) -- (I17) -- (B70) -- (-2,2) -- cycle;
	\addlegendimage{area legend,fill=Turquoise};
	\filldraw[thick,fill=CadetBlue] 
	(B10) -- (B70) -- (I17) -- cycle;
	\addlegendimage{area legend,fill=CadetBlue};
	\filldraw[thick,fill=CornflowerBlue] 
	(B51) -- (B31) -- (I38) -- (B81) -- (B50) -- cycle;
	\addlegendimage{area legend,fill=CornflowerBlue};
	\filldraw[thick,fill=Emerald] 
	(B30) -- (B81) -- (I38) -- cycle;
	\addlegendimage{area legend,fill=Emerald};
	\filldraw[thick,fill=gray!40] 
	(B50) -- (B51) -- (2,2) -- cycle;
	\addlegendimage{area legend,fill=gray!40};
	\legend{$\Ccal_1$,$\Ccal_2$,$\Ccal_3$,$\Ccal_4$,$\Ccal_5$,$\Ccal_6$,
		$\Ccal_7$,$\Ccal_8$,$\Ccal_9$,$\Ccal_{10}$,$\Ccal_{11}$,$\Ccal_{12}$,
		$\Ccal_{13}$,$\Ccal_{14}$,$\Ccal_{15}$,$\Ccal_{16}$};
\end{axis}
\end{tikzpicture}
\subcaption{$\Hcal$ and $\Rcal$}\label{fig:casec}
\end{subfigure}
\end{center}
\caption{Cut-Based PWA approximation of the nonlinear function $\dot{s} = \sin(s+u^2)$ using \blue{local modes in (\ref{fig:casef}) shown in the same color as their corresponding subregion in (\ref{fig:casec}). The hyperplanes are defined by 8 pairs of points on $\Dcal$ to form the cutting arrangement $\Hcal$ (see Fig.~\ref{fig:def2}), resulting in partitioning the domain into 16 subregions $\Rcal$ (see Fig.~\ref{fig:def3}).}}
\label{fig:case}
\end{figure*}

To illustrate our cut-based PWA approximation approach, consider the system 
\blue{\[\dot{s} = \sin{\left(s + u^2\right)},\]}
with a 2-dimensional domain 
\blue{\[\Dcal \coloneqq \{\; (s,u) \in \mathbb{R}^2 \; \; \vert \; \; s \in [-2,2] \; , \; u \in [-2,2] \; \},\]}
to be able to display the function and the subregions. We approximate the nonlinear function \blue{using} Algorithm~\ref{alg:pwa} by selecting the stopping criterion 
$\text{tol}_{\rm{err}} = 5\%$
and imposing the continuity constraint in (\ref{eq:lowlevelc}). The algorithm reaches the stopping criterion with $\nc=8$ cuts by partitioning the domain into $P=16$ subregions. Figure~\ref{fig:case} shows the resulting PWA approximation, region set, and the cutting hyperplanes.

\subsection{Comparison with Approaches in the Literature}

The comparison analysis is done separately for each approach from the literature, using case studies form its respective publication.

\subsubsection{Mesh-based recursive abstractions}

Here, we use the same benchmark incorporated in~\cite{Jin2021}: the Dubins car dynamics \mbox{$F:[-2,2]\times[0,2\pi] \to \mathbb{R}$} given by
\[F(x) = x_1 \cos(x_2).\]
The abstraction error is defined as the distance between the upper and lower PWA approximations of $F$. Therefore, for a fair comparison, we assume the approximation error to be half of the abstraction error. 

Table~\ref{tab:comp} compares the number of required subregions to reach three different user-defined maximum approximation errors. It is evident that the new cut-based approach can reach the same accuracy level with significantly fewer subregions. However, it should be noted that the two methods introduced in~\cite{Jin2021} converge to their minima faster than our approach. Therefore, the benefit of the new cut-based method is best realized for offline computations aimed at reaching a more accurate, yet simpler, PWA approximation forms. 

\begin{table}[htbp]
\caption{Number of required subregions using different PWA approximation methods}
\begin{center}
\begin{tabular}{c|c c c}
	\hline
	 & \multicolumn{3}{c}{\textbf{Maximum error}}\\
	 \textbf{Approach} &	10\% & 5\% & 2.5\% \\
	\hline
	Cut-based approach & 12 & 24 & 40\\
	Method I~\cite{Jin2021} & 58 & 108 & 210\\
	Method II~\cite{Jin2021} & 61 & 121 & 257\\
	\hline
\end{tabular}
\label{tab:comp}
\end{center}
\end{table}

\subsubsection{Parametric plane-based cutting strategy}

In our previous paper~\cite{IFAC2023}, a cut-based PWA approximation is done by pairing the input-state vectors to form 2D hyperplanes and the cutting hyperplanes are defined perpendicular to the selected pairs. As one of the illustrative examples in~\cite{IFAC2023}, the longitudinal velocity of a single-track vehicle model with linear tires is denoted by \mbox{$F : \mathbb{R}^6 \to \mathbb{R}$} is given by
\begin{multline*}
	F(x,u) = \dfrac{u_1 \cos (u_3) + u_2}{\eta_1} + \\
	\eta_2 \left[\tan^{-1} \left(\dfrac{x_2 + \eta_3 x_3}{x_1}\right) - u_3\right],
\end{multline*}
with $\eta_1 = 1970$, $\eta_2 = 64.36$, and $\eta_3 = 1.48$, incorporating 3 states and 3 inputs. To approximate $F$, 3 non-intersecting cuts are required on $(x_2,u_3)$ to partition the domain into 4 subregions and achieving the -- significantly high -- approximation error of 30\%. Moreover, the resulting PWA approximation is not continuous.

Table~\ref{tab:compifac} compares different cut-based PWA approaches in terms of the continuity of their respective number of regions, their maximum error, and the continuity of their PWA form. Using our proposed cutting strategy, we are no longer limited to plane-based cutting of the domain, which significantly improves our flexibility in partitioning the domain, which results in obtaining a PWA approximation with 3 local modes for a maximum error of 3\%. Moreover, we can now enforce a continuous PWA approximation of the dynamics. As a result, we are able to reach 3\% approximation error using only two cuts: one on $v_y-r$ and one on $v_x-v_y$ axis. 

\begin{table}[htbp]
	\caption{Comparing cut-based PWA approximation approaches}
	\begin{center}
		\begin{tabular}{c|c c c}
			\hline
			& \textbf{Continuous} & \textbf{No.\ of} & \textbf{Max.} \\
			\textbf{Approach} & \textbf{from} & \textbf{regions} & \textbf{error} \\
			\hline
			Current method & Yes & 4 & 3\% \\
			Current method & No & 3 & 3\% \\
			Previous method~\cite{IFAC2023} & No & 4 & 30\% \\
			\hline
		\end{tabular}
		\label{tab:compifac}
	\end{center}
\end{table}

\section{Conclusions}\label{sec:conc}

In this paper, we have proposed a novel approach for PWA approximation of nonlinearities using a hinging-hyperplane formulation to partition the domain into subregions. Our proposed method does not require prior knowledge of the dynamics, is applicable to nonlinear systems defined on multi-dimensional domains, and allows for a straightforward formulation of the continuity constraint for the PWA approximation. To avoid unnecessary complexity in the final form, the number of cutting hyperplanes is iteratively increased in case the solutions of the approximation problem are unable to satisfy a user-defined error tolerance. The flexibility of our proposed approach allows for various polytopic subregion definitions and adaptability for different approximation requirements. By comparing the performance of our approach to other state-of-the-art methods from the literature, we have showcased its potential for practical applications in complex, large-scale systems, paving the way for future advancements in nonlinear function approximation and control. \blue{For future work, it is intended to investigate the relationship between the proposed method and the solutions of the original nonlinear dynamics. In particular, potential implications such as trajectory proximity, existence and uniqueness of trajectories, or possible sliding-mode phenomena should be examined.}

\begin{ack}                              
	This research is funded by the Dutch Science Foundation NWO-TTW within the EVOLVE project (no.\ 18484). 
\end{ack}

\bibliographystyle{unsrt}        
\bibliography{Citations}

\end{document}